# Simulated Bars May Be Shorter But Are Not Slower Than Observed: TNG50 *vs.* MaNGA

Neige Frankel,[1,2,3] Annalisa Pillepich,[2] Hans-Walter Rix,[2] Vicente Rodriguez-Gomez,[4] Jason Sanders,[5] Jo Bovy,[3] Juna Kollmeier,[1] Norm Murray,[1] and Ted Mackereth[1,3]

[1]*Canadian Institute for Theoretical Astrophysics, University of Toronto, 60 St. George Street, Toronto, ON M5S 3H8, Canada*
[2]*Max Planck Institute for Astronomy, Königstuhl 17, D-69117 Heidelberg, Germany*
[3]*Department of Astronomy and Astrophysics, University of Toronto, 50 St. George Street, Toronto, ON M5S 3H4, Canada*
[4]*Instituto de Radioastronomía y Astrofísica, Universidad Nacional Autónoma de México, Apdo. Postal 72-3, 58089 Morelia, Mexico*
[5]*Department of Physics and Astronomy, University College London, London WC1E 6BT, UK*

## ABSTRACT

Galactic bars are prominent dynamical structures within disk galaxies whose size, formation time, strength, and pattern speed influence the dynamical evolution of their hosts galaxies. Yet, their formation and evolution in a cosmological context is not well understood, as cosmological simulation studies have been limited by the classic trade off between simulation volume and resolution. Here we analyze barred disk galaxies in the cosmological magneto-hydrodynamical simulation TNG50 and quantitatively compare the distributions of bar size and pattern speed to those from MaNGA observations at $z = 0$. TNG50 galaxies are selected to match the stellar mass and size distributions of observed galaxies, to account for observational selection effects. We find that the high-resolution of TNG50 yields bars with a wide range of pattern speeds (including those with $\geq 40 \, \rm km \, s^{-1} \, kpc^{-1}$) and a mean value of $\sim 36 \, \rm km \, s^{-1} \, kpc$ consistent with observations within $6 \, \rm km \, s^{-1} \, kpc^{-1}$, in contrast with previous lower-resolution cosmological simulations that produced bars that were too slow. We find, however, that bars in TNG50 are on average $\sim 35\%$ shorter than observed, although this discrepancy may partly reflect remaining inconsistencies in the simulation-data comparison. This leads to higher values of $\mathcal{R} = R_{\rm corot}/R_{\rm b}$ in TNG50, but points to simulated bars being 'too short' rather than 'too slow'. After repeating the analysis on the lower-resolution run of the same simulation (with the same physical model), we qualitatively reproduce the results obtained in previous studies: this implies that, along with physical model variations, numerical resolution effects may explain the previously found 'slowness' of simulated bars.

## 1. INTRODUCTION

Bars, linear features in the stellar surface brightness at the center of galaxies, are commonly observed in disk galaxies (e.g. Masters et al. 2011; Erwin 2018). They affect the appearance of their host galaxy, and how they dynamically evolve (e.g., Lynden-Bell & Kalnajs 1972; Weinberg 1985; Athanassoula 2003; Kormendy & Kennicutt 2004; Minchev & Famaey 2010; Sellwood 2014; Chiba et al. 2019; Garma-Oehmichen et al. 2021) . Bars can form via various channels (a) instabilities in self-gravitating stellar disks (e.g., Hohl 1971; Ostriker & Peebles 1973), (b) tidal perturbations by flyby perturbers such as satellite galaxies (Peschken & Łokas 2019; Miwa & Noguchi 1998; Noguchi 1987; Łokas 2021), or (c) a combination of the two (Miwa & Noguchi 1998). After their formation, they evolve by interacting with the rest of their host galaxy, both its stellar (Athanassoula 2003) and dark matter components (Tremaine & Weinberg 1984a). Bars are often assumed to form such that they fill their corotation radius $R_{\rm corot}$, i.e. their size $R_{\rm b}$ extends to the radius in the disk where stars on circular orbits orbit about the galactic center at the bar's angular speed, and that the two main processes involved in their subsequent evolution are (a) dynamical friction,

frankel@cita.utoronto.ca

slowing down the bar with only modest growth - if any[1] (Weinberg 1985; Debattista & Sellwood 2000), and (b) trapping disk stars, leading the bar to both grow and slow down (e.g., Athanassoula 2003).

In observations, we can only measure the detailed properties of the bars (size, pattern speed, strength) for external galaxies at $z \approx 0$ (Font et al. 2017; Guo et al. 2019; Aguerri et al. 2015; Williams et al. 2021) and for the Milky Way (Sanders et al. 2019; Bovy et al. 2019; Hinkel et al. 2020) i.e. after they have formed and evolved until their present-day state. To draw conclusions on their histories requires us to use strong assumptions and to work with distance-independent quantities, such as the ratio of the corotation radius to bar size $\mathcal{R} = R_{\rm corot}/R_{\rm b}$ (Elmegreen et al. 1996). Assuming that bars form with $R_{\rm b} = R_{\rm corot}$ and that dynamical friction and bar growth are the only two processes affecting a bar, a large $\mathcal{R}$ has been interpreted as a bar that slowed down a lot by dynamical friction, and a small $\mathcal{R}$ as a faster bar (with the boundary at $\mathcal{R}= 1.4$). Presuming dark matter absorbs angular momentum via dynamical friction, such reasoning has been typically used to draw conclusions on the inner dark matter content of barred galaxies. However, uncertainties on $R_{\rm b}$ and $R_{\rm corot}$ can be large and the definition of $R_{\rm b}$ affects the resulting $\mathcal{R}$ value, leading to controversial interpretations on whether some observed galaxies have 'ultra-fast'[2] or 'slow' bars (e.g., Cuomo et al. 2021). Therefore, the definition-dependent values taken by $\mathcal{R}$, in the absolute sense, give only limited insights on the evolution of barred galaxies. However, comparisons with simulations using the same bar size definition might help understanding the processes setting the bar properties.

Controlled simulations of improved realism have shown that the picture from the previous paragraph may be too simple. More than only two processes can affect the dynamics of bars: for instance, when galaxies contain gas (Berentzen et al. 2007; Athanassoula et al. 2013; Bi et al. 2021), this could exchange angular momentum, or destabilize the bar (Hasan et al. 1993), the latter being unlikely in realistic systems (Shen & Sellwood 2004). The matter content of a galaxy (and its halo) is not the only important aspect setting bar properties, as their kinematic state also matters (Athanassoula 2003).

In a cosmological context, the properties of bars at formation are not well understood, and bars may not form filling their corotation radius as commonly assumed (Bi et al. 2021). Mergers and satellites may also exchange angular momentum, energy, and mass, and hence may affect the bar properties (Gerin et al. 1990; Ghosh et al. 2021; Bortolas et al. 2020, 2021) in complex ways: by affecting the bar pattern speed (Martinez-Valpuesta et al. 2017; Miwa & Noguchi 1998), by enhancing its strength (e.g., at pericenter passage), or by destroying it (by bringing large mass of ex-situ stars to the central part of the galaxy). These processes in turn can affect the bar fraction, which has been extensively used as a test for cosmological simulations (Algorry et al. 2017; Peschken & Łokas 2019; Zhao et al. 2020; Rosas-Guevara et al. 2020; Reddish et al. 2021; Rosas-Guevara et al. 2021) such as EAGLE (Schaye et al. 2015), Illustris (Vogelsberger et al. 2014), TNG100 (Springel et al. 2018; Pillepich et al. 2018a; Nelson et al. 2018; Naiman et al. 2018; Marinacci et al. 2018), NewHorizon (Dubois et al. 2021), and TNG50 (Pillepich et al. 2019; Nelson et al. 2019a). However, fully-cosmological simulations have so far been reported to produce galactic bars that are too slow[3] (Peschken & Łokas 2019; Algorry et al. 2017; Roshan et al. 2021b), although the observational and simulation galaxy samples put in comparison had different distributions of properties (mass, size, etc.) which could partly account for differences between observations and simulations. State-of-the-art zoom-in simulations such as AURIGA (Grand et al. 2017) and NIHAO (Buck et al. 2020) have more recently produced galaxies with bars that rotate with large pattern speeds and that extend to their corotation radii (Fragkoudi et al. 2020; Hilmi et al. 2020). This suggests that either the numerical resolution or the subgrid physics influence the pattern speed of bars, or both (Fragkoudi et al. 2020).

However, to match the observations and to truly see whether simulations produces realistic bar properties (and their distributions), one requires a large sample of simulated galaxies that were selected in a way similar to the observed galaxies. In an attempt to perform a fair comparison, here we use the TNG50 simulation (Pillepich et al. 2019; Nelson et al. 2019b) that reaches zoom-in-like resolution in a large volume (∼50 Mpc simulation box). We contrast it to data from the SDSS (Albareti et al. 2017) MaNGA (Bundy et al. 2015) survey, where galaxies targeted for spectroscopic observations have a simple selection function (Wake et al. 2017), and where galaxy properties were determined homogeneously (similar observational setup, unique pipeline, same analysis tools). Barred MaNGA galaxies were identified via visual inspection, and their strengths, pat-tern speeds, and sizes were estimated in Guo et al. (2019). In the following, we select galaxies in TNG50 that match the observational properties of the observed sample in the stellar mass - effective radius plane, and compare the distributions of the bar properties.

---

[1] the bar can adjust its shape in reaction to angular momentum loss (Weinberg & Tremaine 1983)

[2] term used when $\mathcal{R} < 1$

[3] in this work, we use 'slow' and 'fast' in the absolute sense, fully set by the pattern speed $\Omega_P$, although the literature cited here mostly use 'slow' for $\mathcal{R} > 1.4$

We present the observed and simulated galaxy data in Section 2. In Section 3, we compare the bar size, pattern speed and $\mathcal{R}$ value distributions in the TNG50 simulation to those in the MaNGA observations. We repeat the analysis with the same simulated volume and physical model, but run at lower resolution in Section 4. We discuss possible origins for (a) the matches and mismatches between TNG50 and observations, and (b) the previous mismatches in previous literature in Section 4, and then summarize the results and give an outlook in Section 5. We include a series of appendices detailing the derivation of the pattern speeds (A), exploring more deeply the effects of both stellar mass estimates (B) and of numerical resolution (C).

## 2. OBSERVED AND SIMULATED BARRED GALAXY SAMPLES

To compare the distributions of bar properties in simulations to those in observations, we must choose an observed galaxy sample where the selection criteria and definitions of relevant quantities can be approximately matched in the simulations. Here we use the observed sample of Guo et al. 2019, thereafter G19, the bar properties they derive, and reproduce these observational measurements and selection with TNG50.

### 2.1. Barred External Galaxies: the MaNGA Sample

#### 2.1.1. Selection

We use the sample of barred galaxies presented in G19, which consists of 53 barred galaxies in the 13th data release of SDSS-IV (Albareti et al. 2017; Blanton et al. 2017) MaNGA (Bundy et al. 2015) survey. The galaxies selected for MaNGA IFU observations satisfy simple selection criteria: cuts in photometric estimates of stellar mass, and galaxy size requirements to match the IFU sizes (Wake et al. 2017).

The barred sample presented in G19 involves further selection steps, in particular one involving a human action in the classification of 'barred' and 'unbarred' from galaxy images in Galaxy Zoo2 (Willett et al. 2013) and additional filtering of galaxies for which bar properties could be measured G19. The total stellar mass-size distribution of these galaxies (barred and main sample) is displayed in the middle panel of Fig. 1.

#### 2.1.2. Data Products

The MaNGA sample of barred galaxies presented in Guo et al. (2019) have several spectro-photometric estimates of stellar mass ($M_\star$) using different methods (e.g., Pace et al. 2019), sizes ($R_{\rm eff}$) in the SDSS-r band, mass-weighted bar pattern speeds ($\Omega_P$) and bar sizes ($R_{\rm b}$). We chose to work with the stellar mass estimate of Pace et al. (2019) in this work, but also show results with different estimates in the Appendix B. Pace et al. (2019) obtained stellar masses by modelling stellar populations in order to match the spatially-resolved observed integrated spectra in the IFU aperture and the photometry of the observations. They use a large library of star formation histories and the associated optical spectra to fit the spectral energy distributions in the MaNGA galaxies. Pace et al. (2019) then derive (1) an estimate of the stellar mass within the IFU aperture, and (2) an estimate of the total stellar mass, correcting for the unseen mass outside of the IFU aperture. For illustrative purpose, we use the total mass (aperture-corrected) in the first two panels of Fig. 1. However, for the quantitative work in Section 2.3 and the right panel of Fig. 1, we use the mass inside of the IFU aperture as the fiducial choice.

After visually inspecting all the galaxies in this sample, we find that for a few galaxies, bar size measurements appear counter intuitive: namely, the quoted bar size in Guo et al. 2019 seems inconsistent with an SDSS-r image of the galaxy, or the galaxy's inclination measurement seems unrealistic. We chose to keep working with these data because we did not identify a clear procedure or algorithm to support discarding them nor a way to improve the measurement. However, since most of the ambiguous systems are those with high inclination measurements, we do plot the high-inclination systems ($i > 60°$) contributions to the distributions of bar properties in a different color, so that they can still be 'separated' when interpreting the resulting figures.

### 2.2. Barred Simulated Galaxies: the TNG50 Sample

#### 2.2.1. The TNG50 simulation

The TNG50 simulation (Nelson et al. 2019b; Pillepich et al. 2019) is a high-resolution cosmological magneto-hydrodynamical (MHD) simulation reaching a zoom-in resolution in a fully cosmological context. It includes the coupled effects of gravity, MHD, star formation, gas cooling and heating, and feedback from stars and supermassive blackholes (Weinberger et al. 2017; Pillepich et al. 2018b). The latter were designed to reproduce global properties of galaxies and galaxy populations (e.g., the $z = 0$ galaxy stellar mass function and stellar-to-halo-mass relation and the star formation rate density across time) but not the details of their inner content such as bar properties, which are then predictive (and that we set out to test here). Stellar feed-back leads galactic outflows via star formation-driven kinetic, decoupled winds. The supermassive black hole feedback works in a thermal 'quasar' ('kinetic wind') mode at high (low) accretion rates. TNG50 reaches zoom-in numerical resolution with a baryonic (dark) mass of $\sim 8.5 \times 10^4 M_\odot$ ($4.5 \times 10^5 M_\odot$) and a cell size of 70-140 pc on average in star-forming regions. The simulation box is large enough (L=50 Mpc) so that at $z = 0$, over 130 galaxies in the stellar mass range $10^{10} M_\odot \leq M_\star \leq 10^{11.5} M_\odot$ have a disk and bar, allowing a statistical study of bars in the simulation.

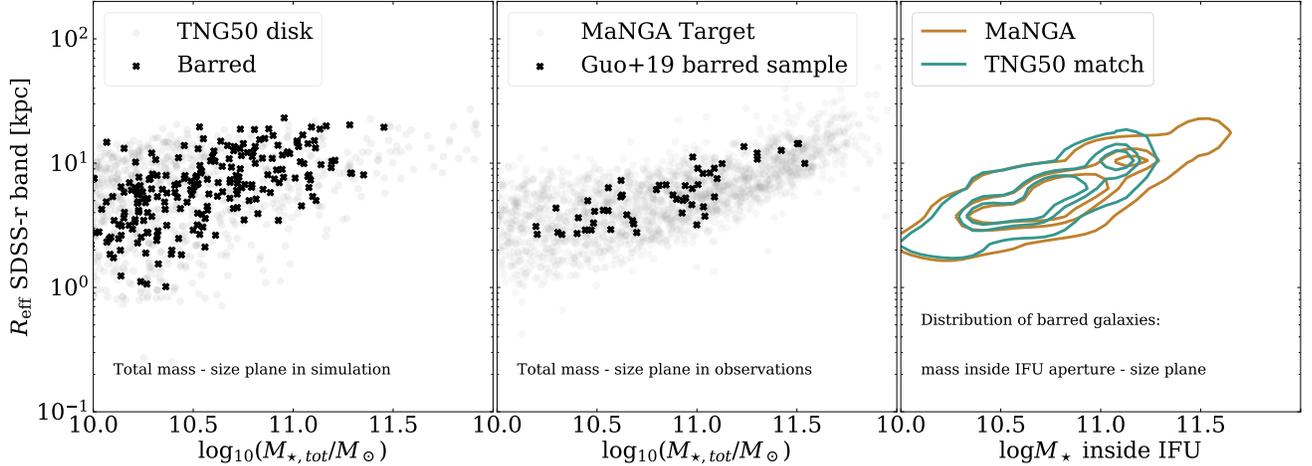

**Figure 1.** Matching the stellar mass-size plane for the simulated galaxies from TNG50 (left) and the observed galaxies from MaNGA (middle). The grey dots represent all disk galaxies (at face value) and the black crosses are the barred galaxies in TNG50 (left), and the G19 sample targeted by MaNGA and identified as barred via visual inspection (middle). Here, total mass means 'all bound stars' in TNG0 and 'total stellar mass estimate' from Pace et al. (2019)' for MaNGA. The third panel shows contours of the distributions of barred galaxies matched between MaNGA and TNG50 in the plane of *stellar mass inside IFU* and effective radius, quantities that were used to match the galaxy samples.

### 2.2.2. Identifying TNG50 barred galaxies

We construct a sample of barred galaxies in the final snapshot ($z = 0$) of the TNG50 simulation by building on the methodology presented in Rosas-Guevara et al. (2020). We project each galaxy face-on with the stellar angular momentum vector pointing to increasing $z$; we select disk galaxies, i.e. supported by rotation, by imposing that over 40% of the stars have circularities $\geq 0.7$. We measure the strength of the Fourier modes of the stellar surface density $A_m(R)$ for $m = (0, 2, 4, 6)$ as a function of galactocentric radius, and define the $m = 2$ strength, $A_{2\mathrm{max}}$, as $A_{2\mathrm{max}} = \max(A_2(R)/A_0(R) \mid R < R_{max,def})$, where $R_{max,def}$ is a visually imposed maximum radius of the bar extent (this cut proved necessary to impose for some ambiguous cases like spiral galaxies). We consider galaxies where $A_{2\mathrm{max}} > 0.2$ to be barred, and utilize their present-day properties in the remainder of this work.

### 2.3. Selecting and Matching Galaxies for Comparisons

To quantitatively compare the simulations to the observations, it is crucial for the two datasets to undergo the same selection. As mentioned above, some selection criteria are approximately reproducible (stellar mass, size), but human decision cannot be reproduced (specifically the bar classification). To approximately bring the observed and simulated sample together, we match the TNG50 galaxies to the properties *that enter the selection of the observed galaxies* (i.e. only mass and size). For each observed barred galaxy $j_\mathrm{obs}$, we find the 5 nearest TNG50 barred galaxies in the stellar mass - effective radius plane, irrespective of whether they are central or satellite. We could employ a different matching method but find that this procedure reproduces sufficiently well the distributions in the mass-radius plane.

*Stellar mass.* To minimize the possible systematics in the model of the stellar mass outside of the IFU in Pace et al. (2019), we only use and measure the stellar mass inside of an aperture of the same size (IFU size) in TNG50 galaxies (and test two other stellar-mass matching procedures in Appendix B). The stellar mass enclosed in an IFU aperture $R_\mathrm{IFU}$ is:

$$M_{\star,\mathrm{IFU}} = \sum_i m_{\star,i} b(R_i) \qquad (1)$$

where the sum is taken over the masses of stellar particles, and the binning function is such that

$$b(R) = \begin{cases} 1 & \text{if } 0 < R < R_{\star,\mathrm{IFU}}^{j_\mathrm{obs}} \\ 0 & \text{otherwise.} \end{cases} \qquad (2)$$

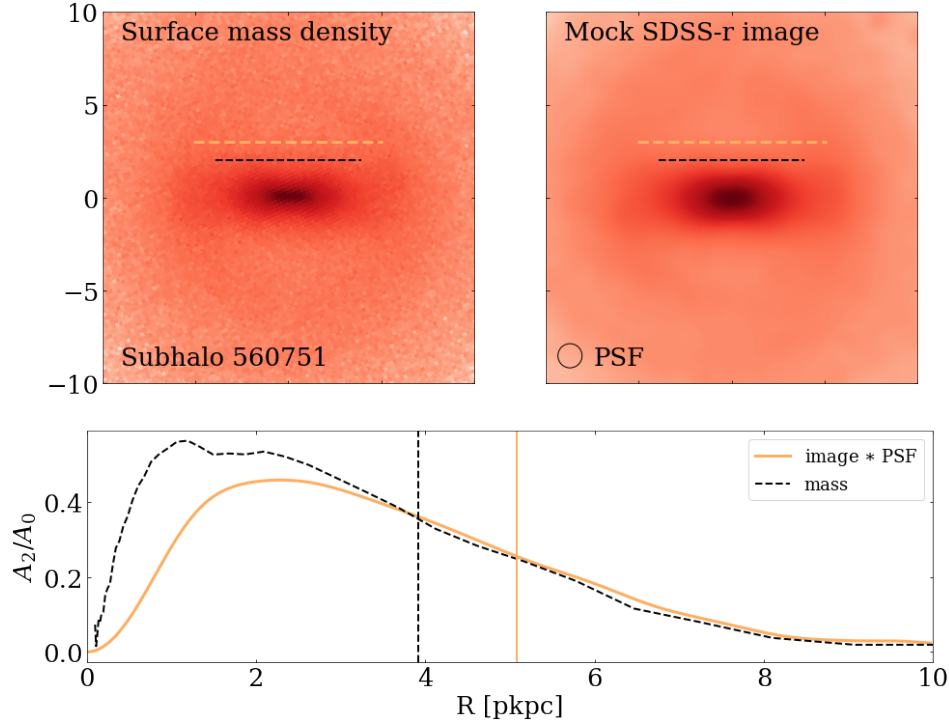

**Figure 2.** Illustrating the determination of the bar length for an example TNG50 galaxy at $z = 0$. Top left: Surface density map of the galaxy (representing the real stellar mass distribution). Top right: face-on image of the galaxy constructed with SKIRT (Baes et al. 2011; Camps & Baes 2015; Rodriguez-Gomez et al. 2019) with an SDSS-r band PSF of FWHM=1.3 arcsec. The horizontal dashed lines show the bar extent derived from the mass (black) and the PSF-convolved mock image (orange). Bottom: radial profile of the normalized Fourier component ($A_2/A_0$) for the mass distribution (dashed black) and the mock image with PSF convolution (orange). The vertical lines represent the bar length as measured from the mass and PSF-convolved light profiles respectively.

The IFU aperture is calculated in physical size for each observed galaxy from the number of fibers used to observed each galaxy, the fiber size, and the distance to each galaxy. The diameter of a single fiber is $\approx 2''$ (Drory et al. 2015). The diameter of the full aperture goes as $\sqrt{N_{\rm fibers}}$, where $N_{\rm fibers}$ is the number of fibers, and its relation to the total diameter of a field is a documented data product (Law et al. 2016). As observed galaxies can be at different distances and be observed with different apertures, the mass inside the IFU aperture is recalculated for each TNG50 barred galaxy every time a comparison with a new observed galaxy $j_{\rm obs}$ is made, such that a single TNG50 galaxy can have several different values for its mass inside IFU. The IFU apertures were designed to cover at least 1.5 and 2.5 Petrosian radii (Petrosian 1976) of (most) galaxies. Therefore, the difference between the total stellar mass of a galaxy and its mass inside the IFU aperture is very small.

*Stellar size.* To match observed and simulated galaxies based on stellar sizes, we use for TNG50 galaxies the circularized half-light radii in the SDSS-r band from face-on projections (Pillepich et al. 2019). The stellar light is forward modeled by summing the emission from single stellar populations of given age, initial mass (assumed to come from a Chabrier distribution) and metallicity (Vogelsberger et al. 2013). Here the effects of dust are neglected for the matching, even though it could have non-negligible effects in the determination of galaxy sizes (Nelson et al. 2018; de Graaff et al. 2021).

The matching procedure described above produces a sample of 79 unique simulated galaxies from TNG50 that have a similar mass-size distribution as the observed sample used to make the comparison of the bar properties: see Fig. 1. This approach ensures such comparison to be meaningful, or at least not to be affected by systematic trends of bar properties with other galaxy properties. A single TNG50 galaxy can be matched with several MaNGA galaxies but do not induce a full degeneracy: matching parameters differ on observations (size of the IFU radius to calculate the mass enclosed within that radius) and distance, affecting both the selection and later-on the bar size measurement.

### 2.4. *Defining and Quantifying Bar Properties*

With the above match between observed external galaxies to those in TNG50, we define a set of properties of the bars that are measurable from the simulations and that are analogous to those in observations.

- The **bar strength**, $\mathbf{A_{2max}}$, is defined in Section 2.2 as the maximum of the normalized $m = 2$ Fourier mode in the *mass distribution*, $A_{2max} = \max(A_2(R)/A_0(R))$.

- The **bar size**, $\mathbf{R_b}$, follows G19's definition for a straightforward comparison with observations. It is the radius where

$$\frac{I_b(R_b)}{I_{ib}(R_b)} = 0.5 \times \left[\left(\frac{I_b}{I_{ib}}\right)_{max} - \left(\frac{I_b}{I_{ib}}\right)_{min}\right] + \left(\frac{I_b}{I_{ib}}\right)_{min}, \quad (3)$$

where the bar amplitude is $I_b = A_0 + A_2 + A_4 + A_6$ and the inter-bar amplitude is $I_{ib} = A_0 - A_2 + A_4 - A_6$, where $A_m$ are the Fourier amplitudes of the $m$ modes in the *light distribution*, detailed in Section 2.5.

- The **bar pattern speed**, $\mathbf{\Omega_P}$, is the rate of change of the phase of the $m = 2$ Fourier mode. We obtain it from individual snapshots by using the continuity equation on the mass density and velocity field. This method is conceptually similar to the Tremaine & Weinberg (1984b) method and is based on the same principle, except that we keep the differential form (instead of integrating along an arbitrary line of sight). The derivation is in Appendix A.

- The **bar corotation radius**, $\mathbf{R_{corot}}$, is the radius at which a star on a circular orbit has the same angular velocity as the bar, i.e. the radius where $\Omega_P = \Omega_{circ}(R)$. We measure the circular velocity $v_{circ}(R) = \sqrt{GM(R)/R}$ from the spherically-averaged total enclosed mass inside a radius $R$. This definition is different from that for the observed MaNGA galaxies, for which $v_{circ}$ was determined using Jeans Anisotropic Modelling and velocity maps obtained from the integrated field units spectra (Cappellari 2008). We have also measured the circular velocity curve from a fit of the midplane potential (with AGAMA Vasiliev 2019) given the matter distribution in the simulation and find no significant difference in the main results of this work.

- The $\mathcal{R}$ **value** is the dimensionless ratio of the corotation radius to the bar extent, $\mathcal{R} = R_{corot}/R_b$.

### 2.5. *Light-weighted Bar Sizes*

We obtain the light-weighted bar sizes anticipated in Section 2.4 after emulating SDSS-like synthetic face-on images of TNG50 galaxies following the methodology described in Rodriguez-Gomez et al. (2019). In summary, old stellar particles are assumed to be coeval stellar populations with spectra given by Bruzual & Charlot (2003) and young stellar particles are assumed to be star-forming regions with spectra models from Groves et al. (2008). The effects of dust attenuation and scattering are modeled with SKIRT (Baes et al. 2011; Camps & Baes 2015). Since there is no dust in the simulation, star-forming gas is used as a proxy for dust, assuming a dust-to-metals ratio 0f 0.3. Rodriguez-Gomez et al. (2019) foound that the optical morphologies of IllustrisTNG galaxies are within one standard deviation of their observational counterpart in PanSTARRS. So, whereas the identification of TNG50 barred galaxies is made based on the stellar mass distribution, the bar sizes are measured from the simulated galaxy mock images. By placing the simulated galaxies at the redshift of the observed galaxies they are matched with (of median $z = 0.032$ and spread 0.02), we convolve the images in the SDSS-r band with a PSF of 1.3 arsec FWHM. An example of an image is shown in the top right panel of Fig. 2, where the top left panel is the equivalent in surface mass density at face value from the simulation. The bottom panel shows the strength of the normalized Fourier $m = 2$ mode $(A_2/A_0)$ obtained from (1) the surface stellar mass density, and (2) the PSF-convolved light profile emulating the observations.

Equation (3) can have multiple solutions, in particular for galaxies with strong spirals. For the purpose of a fair comparison with G19, we follow their procedure and visually inspect the galaxy images choosing the solution of the bar size that seems closest to the visual extent of the bar[4]. As a result of the restriction to the r-band, the pixel size, PSF, and dust extinction, the difference between the mass-weighted and light-weighted $A_2$ profiles varies from galaxy to galaxy. On average, the light-weighted bar size estimate is larger by 15% to 20% than the mass-weighted estimate, across the entire stellar mass range.

Fig. 3 shows the TNG50 barred galaxies used in this work and mocked as if observed in SDSS.

## 3. COMPARISONS OF OBSERVED VS SIMULATED BAR PROPERTIES

We now have a sample of barred galaxies in TNG50 that have a similar distribution of the stellar mass and effective radius as the observed galaxies. We evaluate bar sizes, pattern speeds, and corotation radii as described in Section 2.4 and compare their distributions in TNG50 to those in MaNGA. This comparison is shown in Fig. 5.

---

[4] Private communication with Rui Guo clarifying the procedure used in G19.

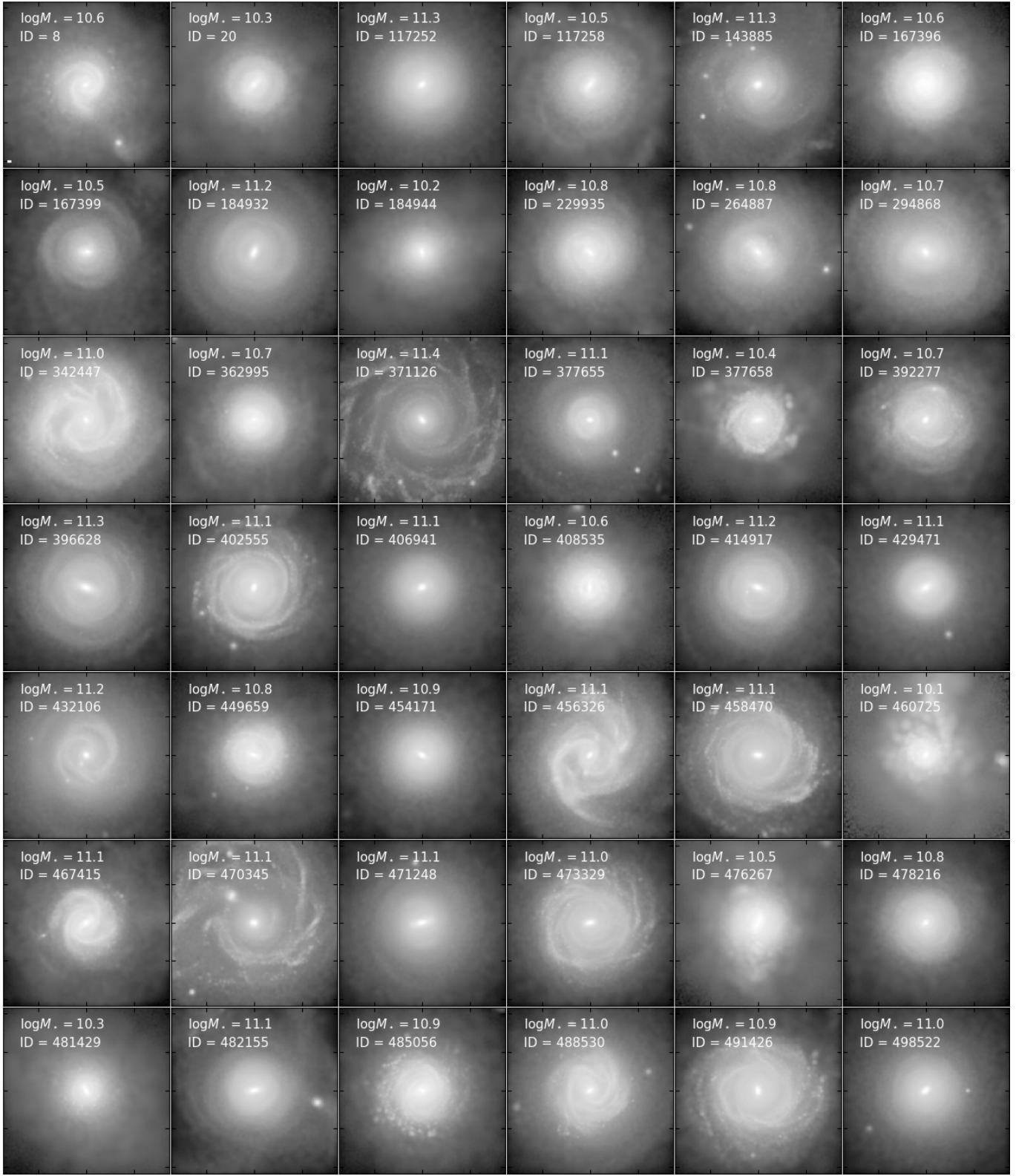

**Figure 3.** SDSS-r band image stamps of the TNG50 simulated galaxies that are matched to MaNGA observed galaxies and are used in this analysis. They are seen face on and are annotated by their subhalo IDs and their log stellar mass/$M_\odot$. In the top left panel, a white horizontal segment gives the scale of 1 kpc.

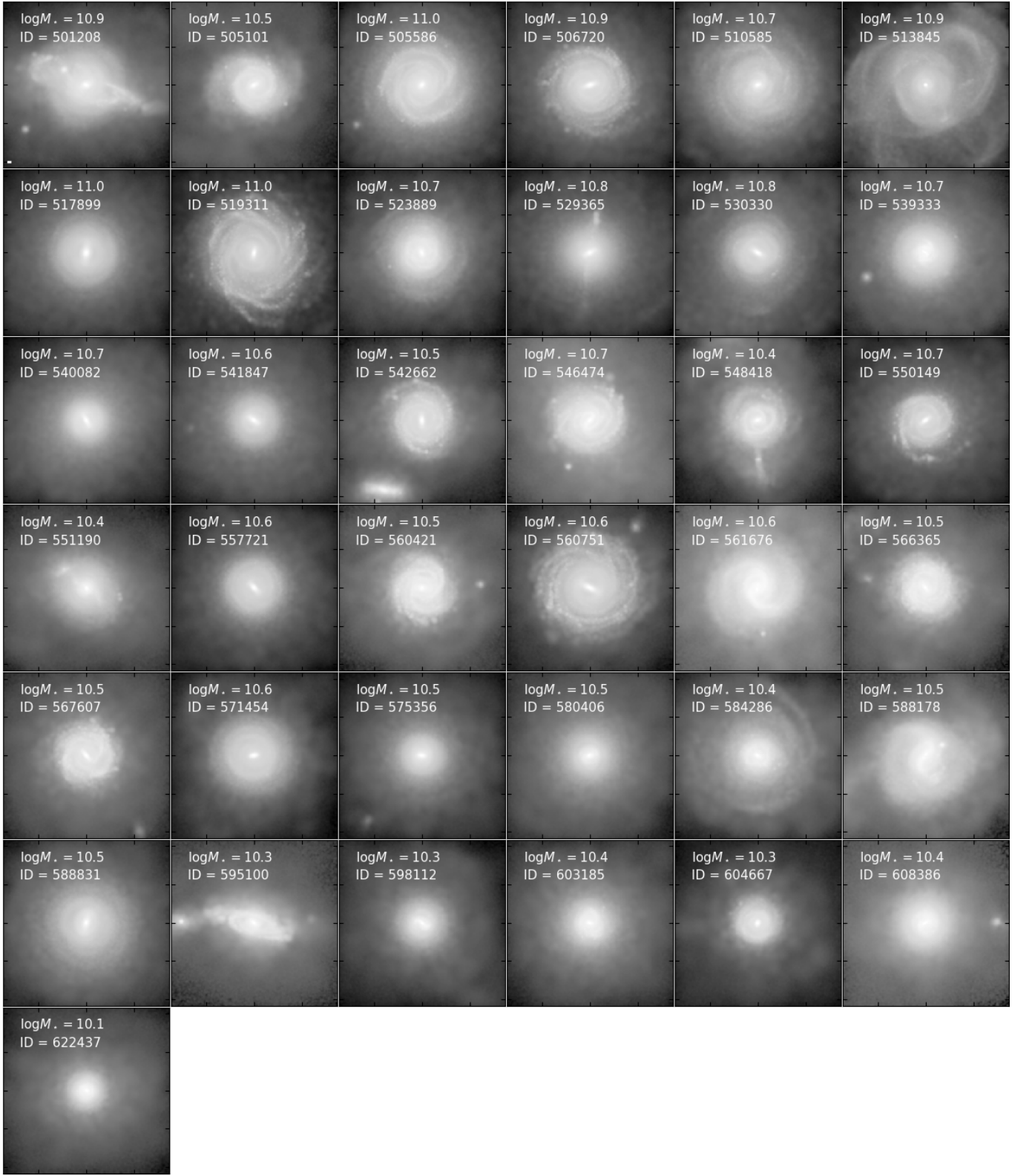

**Figure 4.** Fig. 3 continued.

The pattern speeds of TNG50 simulated galaxies cover the range of those observed well, including pattern speeds as large as $\gtrsim 40$ km s$^{-1}$ kpc$^{-1}$. However, TNG50 pattern speeds are 15% higher on average with $\langle \Omega_P^{\mathrm{TNG}} \rangle \approx 1.17 \langle \Omega_P^{\mathrm{MaNGA}} \rangle$. We discuss in Section 5 how this agreement could be affected by the mass matching: pattern speeds have a small correlation with stellar mass, so the histogram could be displaced left or right depending on the definition of stellar mass adopted for matching.

Similarly, the corotation radii seem to be drawn from similar distributions (see Fig. 5, second panel). The better agreement between corotation radii than pattern speeds indicates that the inner mass profile in the simulations is different from that in the observations.

In fact, the bar sizes in the simulation are distinctly smaller than those in the observations by a factor of $\sim 1.5$ in the mean (Fig. 5, third panel). This significant mismatch could have several origins. Either the detailed selection of galaxies as barred is inconsistent, as it uses different methods in simulations and observations (algorithmic in the simulation, human in the observations), or the simulation physics could over-produce short bars (see Section 5).

As immediate consequence of the three previous comparisons, the $\mathcal{R}$ values are also larger in TNG50 than in MaNGA observations (Fig.5, rightmost panel). Qualitatively, this has been found for the EAGLE and Illustris simulations in previous studies (e.g., Algorry et al. 2017; Peschken & Łokas 2019). We also find qualitatively similar pattern speeds as Roshan et al. (2021b), who also analysed barred galaxies in TNG50, even if with a different galaxy selection. But, compared to that work, we find larger bar size measurements, as these are affected by the galaxy selection, the conversion of mass to light, and the bar size definition. However and importantly, here we claim that the larger $\mathcal{R}$ values in TNG50 galaxies than in MaNGA arise not because the simulated bars are slower, but rather because they are smaller. We note that the definition of the bar size strongly influences the values obtained for $\mathcal{R}$, and a definition of the bar size more closely related to bar orbits would need to be done to interpret the exact values of $\mathcal{R}$ in terms of bar evolution processes.

## 4. EFFECTS OF NUMERICAL RESOLUTION

To investigate whether the numerical resolution of the simulation plays a role in the distributions of the bar properties, we repeat the analysis with the lower-resolution run of the TNG50 simulation (TNG50-2): TNG50-2 has a mass (spatial) resolution worse than the high-resolution run TNG50 (aka TNG50-1) by a factor of 8 (2).

We repeat the matching and measurement procedure described in Section 2: we match TNG50-2 barred galaxies to the mass-size plane of the observed sample, produce SDSS-r band mock images, and measure the bar sizes and pattern speeds. We find that, based on the same selection criteria, the pattern speeds of the bars in the lower-resolution run TNG50-2 are on average lower than those in the observed galaxies by a factor 2: see Fig. 6. Moreover, the pattern speeds of the lower-resolution bars in TNG50-2 do not reach as high values as TNG50 and observed galaxies. The corotation radius distribution differs significantly from that of the observed galaxies. However, the bar sizes have a similar distribution to the MaNGA observations (excluding again the very long observed bars), and the average $\mathcal{R}$ value is higher in the simulation by a factor 2. Qualitatively similar conclusions would be derived if, instead of directly matching TNG50-2 galaxies to the observed MaNGA sample, we inspect the bar properties of the low-resolution counterparts of the TNG50 galaxies used throughout this analysis (see Section C.1).

Our results are qualitatively in line with previous literature results based on the previous generations of cosmological simulations, such as EAGLE and Illustris (e.g., Algorry et al. 2017; Peschken & Łokas 2019; Roshan et al. 2021b), which have a similar (or inferior) resolution than TNG50-2: at low resolution, bars are too slow, both in the physical sense (low pattern speed), and in the $\mathcal{R}$ sense (they do not fill their corotation radius).

However, the conclusion we come to here is that numerical resolution plays a central role in determining bar properties in $z \sim 0$ galaxies. Here, we (at least partly) attribute the $\mathcal{R}$-value discrepancy between previous, low-resolution simulations and observations to their numerical resolution being too low and, in fact, too low to fully hold the physical choices responsible, as instead previously claimed (Fragkoudi et al. 2020; Roshan et al. 2021b), either in terms of baryonic physics or of dark matter model.

Additional arguments in this direction and to the effects of numerical resolution on simulated bar properties are given in Appendix C: there, we show that a longer gravitational softening length decreases the forces (and so rotation) and may lead to under-dense inner regions (Section C.3). We also show that, within the galaxy-formation model and resolution of TNG50, the dominance of baryons within the inner regions of galaxies is only mildly lower in the low-resolution run and correlates only weakly with $\mathcal{R}$ (Section C.4). However, at least in the case of the physical and numerical model of TNG50, even though bar properties may be converg*ing*, we cannot say to what degree they are (non-)converg*ed* at the resolution of TNG50 (as we could do with galaxy properties more generally, Pillepich et al. 2019). This is not possible because an even higher-resolution run would be required, as the lower-resolution runs that are available (e.g. TNG50-3 and TNG50-4) are too coarse to even allow us to extract bar properties. These issues complicate the ultimate assessment as to how bar properties also depend on aspects of the physical models (Fragkoudi et al. 2020; Roshan et al. 2021a).

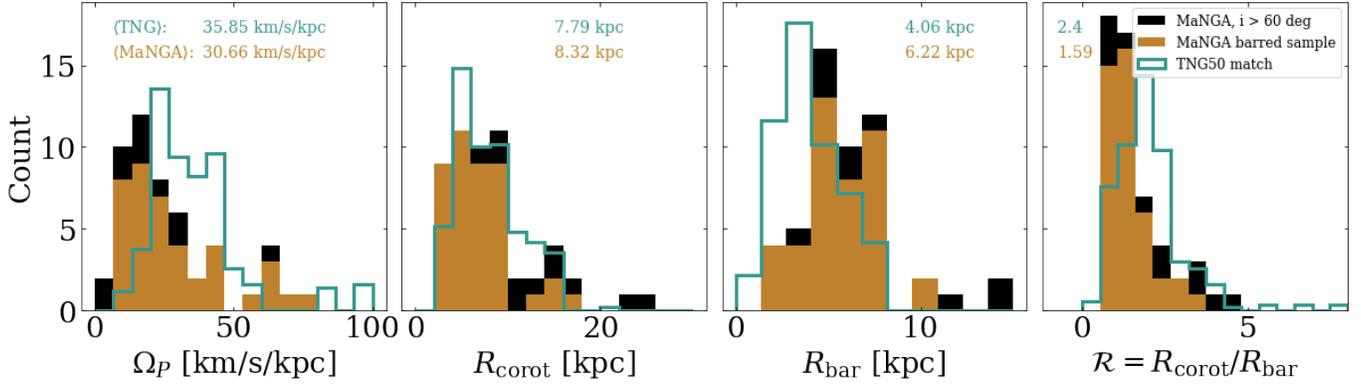

**Figure 5.** Comparing bar properties between TNG50 and MaNGA observations with the distributions of pattern speeds, corotation radii, bar sizes derived from the light, and $\mathcal{R} = R_{\rm corot}/R_{\rm b}$ values for observed and simulated barred galaxies. These are from the MaNGA (filled brown) and the TNG50 (green line) matched samples. Contributions from the highly-inclined MaNGA systems ($i > 60$ deg) are separated out in black. TNG50-obtained bar sizes are emulated to observations. The mean values of pattern speeds, $\Omega_P$, and corotation radii, $R_{\rm corot}$, agree within 15% whereas the mean bar sizes, $R_{\rm b}$, and $\mathcal{R}$ value differ by $\sim 50$%. The averages are annotated at the top of each panel for the simulations and observations respectively. Averages over observed properties also include the highly inclined systems.

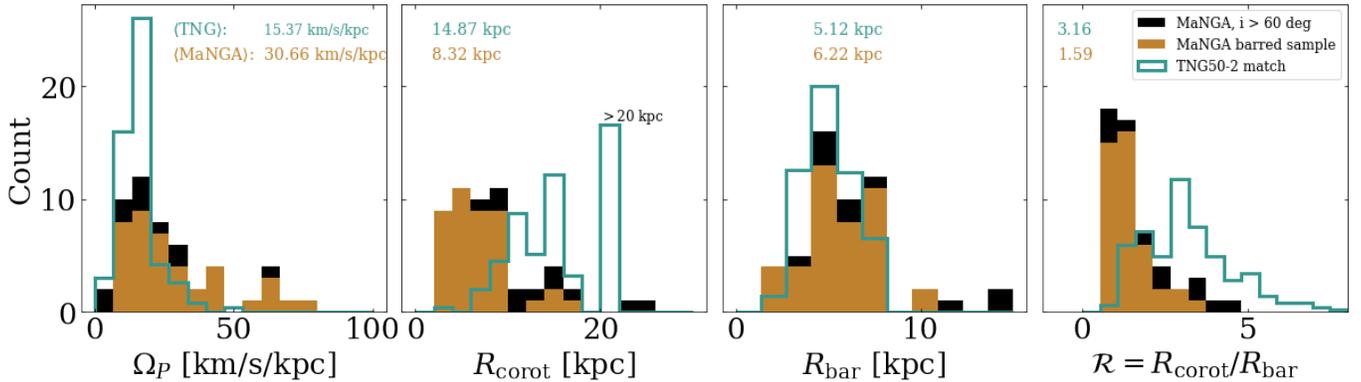

**Figure 6.** Same as Fig. 5, with the analysis (including the matching to MaNGA) repeated with a lower-resolution version of the TNG50 simulation: TNG50-2. The pattern speed in the lower-resolution simulation is lower than that in the high-resolution run by a factor $\sim 2$ and compares poorly with the observed pattern speeds, with no high pattern-speed values. The bar sizes in the lower-resolution run match those observed well. This combination of lower pattern speed and comparable bar size leads to large $\mathcal{R}$ values in the simulation, as seen in previous generation cosmological simulations..

## 5. DISCUSSION

### 5.1. *Taking Observations and Simulations on the Same Ground*

TNG50 is the first cosmological simulation reaching zoom-in resolutions with a volume that is large enough to allow us to quantitatively compare the distributions of simulated galactic bar properties with observations. In this study, we have worked towards a consistent data-model comparison in three ways. We have: 1) sampled the distributions of galaxy properties that enter the selection functions of the surveys, to reduce systematic biases between samples before comparing them; 2) emulated images of the simulated galaxies in the same observed band and at the same angular resolution as in the observations, for consistent bar size measurements; and 3) used the same bar size definition as in the observations. To our knowledge, this is the first time that these steps have been accounted for in a systematic manner for the comparison of simulated and observed bar properties. In fact, these three steps are important for a number of reasons. Firstly, the face-value distributions of galaxy and bar properties in observed samples are, by construction, biased. For example, massive galaxies are brighter and might be over-represented in observations, whereas the stellar mass function in a cosmological simulation box contains more low-mass galaxies. Since on average, low-mass galaxies host bars with lower pattern speeds, comparing simulations to observations without carefully accounting for this effect could lead to the conclusions that simulations produce too slow bars on average, without it being necessarily the case. Secondly, in this paper we have found that measuring the bar size from the mass vs. light profiles affects the measurement by 15%, with light measurements exceeding mass measurements at all stellar masses. Not doing a fair comparison could lead to the (erroneous) conclusion that simulated bars are too short (and in turn too slow, in the $\mathcal{R}$ sense) without it being necessarily the case. In this work, we have accounted for these effects and still find that bars are shorter in the simulation by a factor ∼1.5.

We find that TNG50 bars' corotation radii distribution matches reasonably well that of the galaxies in the MaNGA sample, suggesting that the bars rotate at physically plausible speeds. This is an important step in understanding the dynamical secular evolution of disk galaxies in a cosmological context, as the pattern speeds are crucial in determining the effects of dynamical resonances. This means that the resonances in simulated disk galaxies should be at plausible places in the galaxies, so we can now use them to study the dynamical evolution of TNG50 disk galaxies and expect their physical effects to be plausible. However, since we also find that on average, TNG50 bars are smaller than those in observations, it could weaken the effects of resonances to the outer disk of galaxies. Possible factors that could lead to shorter bars in the simulations are discussed below.

### 5.2. Bar Size Disagreements with Observations

On average, bar sizes in TNG50 are smaller than those in the observations by ∼35%, and this is what drives the $\mathcal{R}$ values to be larger. If this difference is physical, it can serve as a strong constraint for the physical choices made in cosmological simulations. We qualitatively discuss possible origins for this discrepancy. We first list possible inconsistencies in the data-model comparisons which are expanded in Section 5.3, and then discuss its possible physical origins.

Firstly, despite our efforts in making a fair comparison, we did not exactly reproduce the procedure undergone by the data in the observations. Observations do not provide a direct view onto the physical systems of interest. Forward modelling the physical systems all the way down to observed quantities relies on simplifying choices and modeling assumptions in the mass-to-light conversion. In particular:

1. The classification of barred galaxies, which were made by humans on galaxy images for the observed sample, could be biased towards longer bars because long bars are easier to see or are less ambiguous on images, and are better resolved than shorter bars. Erwin (2018) showed that bars shorter than ∼ 2.5 kpc are not resolved in SDSS, whereas our selection in the simulation was purely based on the strength of the Fourier m=2 mode. We show two example of short bars in TNG50 that would probably not be detected with the observational methods in Fig. 13. If we exclude, a posteriori, from the TNG50 matched sample those galaxies whose bars are shorter than 2.5 kpc (and reweight the galaxy matches accordingly), the average values of the pattern speeds, corotation radii, bar sizes and $\mathcal{R}$ values would then agree with the observations within 8%, 2%, 23% and 15% respectively, reducing the bar size discrepancy by a factor ∼ 1.5 and that in $\mathcal{R}$ value by a factor 3. Additionally, G19 report a correlation between bar strength and bar size (longer bars are stronger), which we also find in TNG50. If this correlation is also present at the lower end (which is suggested by the TNG50 bars), and if strong bars are easier to detect, this could imply another bias towards longer bars in the observed sample.

2. We have not simulated images for galaxies with the same inclinations as the observed galaxies. Firstly, the classification of an observed galaxy as 'barred' may depend on its inclination and the presence of dust can affect both the detectability of bars and their size measurement. We find that including dust in the radiative transfer model doubles the number of systems with bar $A_{2max} \leq 0.1$ by a factor two (see a couple of examples in Appendix D), and we would expect this effect to be stronger in inclined systems. On average, we find this effect to be only slightly stronger on short bars. Secondly, the procedure of determining a galaxy's inclination, position angle, and bar size in the inclined image is a complex procedure that can propagate errors and biases, possibly increasing the bar size measurement and the other properties of the galaxies. Since the observed systems with highest inclination measurements did not seem robust (visually checking the presence of a bar or the bar size measurement), we highlighted their contributions to the distributions of observed bar properties in black in Fig. 5. Excluding these systems would decrease the mean bar size of the observed sample, but also reduce the mean corotation radius significantly, bringing the pattern speeds, corotation radii, bar sizes and $\mathcal{R}$ values to differ from the simulations by 7%, 15%, 28% and 75% respectively, increasing the discrepancy between the simulations and observations in $\mathcal{R}$ values.

3. The modeling of the galaxies' light profiles relies on many assumptions. The least robust aspects of the conversion from mass to light are the young stars and the dust, which are sensitive to resolution and model uncertainties, whereas the light produced by the old stellar populations is better known. Since the latter is the dominant component of the stellar bars, it seems unlikely that the discrepancy between the bar sizes in the simulation and in the observations comes from the conversion from stellar particles to light. We have tested (although do not show) the dependence on the bar size measurement on the dust model (by producing images without dust) and find that it would affect the mean bar size of TNG50 galaxies by $\leq 2\%$ at most.

4. The measurement of the effective radius of galaxies, used for the sample matching, was performed differently in the simulation and in the observations (1D radius enclosing half of the light in a face-on projection VS 2D petrosian radius at any inclination). As the bar size can correlate with galaxy size (Erwin 2018; Rosas-Guevara et al. 2020), this inconsistency in the matching procedure could lead to discrepancies in the final bar size comparison.

In the opposite direction, the bar size estimates in TNG50 could be biased towards smaller bars: this may arise because galaxies with two bars generally have the strongest $A_2$ mode from the shorter bar, so we estimate the size of the shorter bar rather than of the long one. Unfortunately it seems practically difficult to reconcile these effects quantitatively within the scope of this work as they involve much human intuition (even if this could in principle be quantified by introducing a Galaxy Zoo project on TNG50 or an algorithmic selection in the observations – see more on this in Section 5.3). Therefore, based on the current information, we cannot tell with certainty whether TNG50 bars are too short or if this is only an effect of the two biases discussed above. Since the lower-resolution run produces, on average, longer bars than the high-resolution run (see sections 4 and C), it seems implausible that the still limited numerical resolution of TNG50 is responsible for the residual shortness of its bars.

Physically, if TNG50 bars are indeed too short, this could arise from several effects that set the inner density profile or prevent bar growth. For example, at fixed numerical resolution, the model for the AGN feedback can strongly influence the central stellar distribution of massive galaxies (Irodotou et al. 2021). An AGN feedback that is too weak can increase the formation of stars at the center of galaxies, producing more compact density profiles and thus shorter, stronger bars. However, there are limitations in tuning the AGN feedback only since it also sets (together with other processes) the overall baryonic mass of galaxies (and it does change it by several factors in Irodotou et al. 2021). This may indicate that if any tuning in the AGN physics were required, then appropriate tuning of other covariant physics may be required too (e.g., star formation, stellar feedback, gas phases etc.). At fixed numerical resolution, varying the physical model on a zoom-in galaxy from the AURIGA to the TNG50 model affects the bar properties (Fragkoudi et al., in prep), with the TNG50 model producing a shorter bar. The details of these specific aspects are extremely complex and covariant and not the focus of the present study, but comparing bar properties to those in the observations may be of help in the physical choices that enter the next generation simulations, and offer a new perspective to understand small-scale physics from large-scale galaxy properties.

The combination of shorter bar in TNG50 and pattern speeds that are more similar to those in observed galaxies than previous cosmological simulations, implies that on average, TNG50 still yields bars with large $\mathcal{R}$ values. Several aspects affecting the formation and subsequent evolution of the bar could lead to this result and would require a more thorough analysis at higher redshifts.

A large $\mathcal{R}$ has generally been interpreted as that the bar slowed down too much due to dynamical friction, as opposed to growing by trapping stars. Such a mechanism could be produced, for example, if the dark matter model in the simulation does not reflect that in Nature and leads to too much dynamical friction (Roshan et al. 2021a). However, if this were the scenario that leads to low $\mathcal{R}$ values here, it would mean that bars were initially born with pattern speeds larger than those in Nature (since they reach the observed present-day pattern speeds).

A bar that is prevented from growing in size due different mechanisms could also be found too short at $z = 0$. For example, keeping disk stars on resonances require them to be on cold orbits. To achieve this in the simulation, high-resolution may be required (making the potential smoother). The presence of other agents such as gas could apply additional effects. Varying the model for the gas and its different phases has been shown to produce structures that clump on different spatial scales and form stars differently (Marinacci et al. 2019), which might affect the bar growth.

5.3. *Limitations of the Simulation-Data Comparison*

In this work we have made efforts towards a data-model comparison that is both rigorous and straightforward. Still, this comparison remains imperfect in a few respects, which we re-iterate and summarize here.

First, we projected all simulated galaxies face-on before calculating their 'observed' properties. If there are biases due to inclination, and since the inclination distribution in the observed MaNGA sample is not uniform (there is an optimum inclination range to be able to measure both pattern speeds with the Tremaine-Weinberg method and bar sizes correctly), then we will not have reproduced this bias. This can be important in measuring galaxy sizes (3D extinction can change the size estimate).

Second, our selection of barred galaxies in the simulation differs from that in the observations. Our selection is purely algorithmic, selecting galaxies that have an $A_2$ strength greater than 0.2, whereas the selection in MaNGA were based on GalaxyZoo results and further visual checks by G19. This limitation could be resolved in the future either with a GalaxyZoo project with Illustris TNG (such as in Dickinson et al. 2018) that would classify barred galaxies in the same way as SDSS observations, or by adopting an algorithm for the classification of barred galaxies in the observations that can be reproduced straightforwardly with simulated data.

We estimated pattern speeds from the 2D velocity and 2D position, rather than being integrated over the line of sight in an arbitrary slit, for some inclination, at some position angle. However, the Tremaine-Weinberg method was tested and found robust in the observed sample by G19, which gives us confidence the results should not be affected significantly.

We compared only to a single type of observations, this could be extended to e.g. the CALIFA samples of barred galaxies, although the selection function in CALIFA is more complex than that in MaNGA, hence our original choice to consider galaxies from the MaNGA survey.

We would need an even higher resolution run to address the question of convergence. The TNG50 simulation suite has an even lower resolution run (TNG50-3) that could in principle be used to see how the differences in bar properties diminish with improving resolution. However, TNG50-3 has a softening length of the same order as typical bar sizes, so taking this approach would not be meaningful (at least for low-mass and small galaxies). Since convergence may also be a function of the physical model used, we also do not make comparisons with different simulations at higher resolutions, and leave this question to be addressed more thoroughly in the future.

In addition to the possible biases in the (non-reproducible) classification of barred galaxies, the at-face value of $\mathcal{R}$ should itself not always be interpreted strictly. The theoretical arguments leading to $\mathcal{R} > 1$ (e.g., Contopoulos & Papayannopoulos 1980) were considering the stability of orbits that align with the bar phase, whereas the $\mathcal{R}$ values obtained from observations do, most of the time, measure bar lengths based on density profiles. Cuomo et al. (2021) recently showed how changing the bar size definition in observations (from light profile analysis to reconstruction of tangential to radial forces) could solve the 'ultrafast bars' observed with $\mathcal{R} < 1$: the bar sizes of these galaxies had been previously over-estimated.

6. CONCLUSIONS

With the combination of high numerical resolution and sample size, the cosmological galaxy-formation simulation TNG50 permits a quantitative comparison of the bar property distributions to observations. In this paper, we have compared MaNGA barred galaxies to a selection of barred galaxies from the TNG50 simulation that are matched in galaxy stellar mass and size distributions. We have measured bar properties from the simulated galaxies as closely as possible to what is done with the observed data, including measuring bar sizes from r-band SDSS-mocked images. Our main results can be summarized as follows.

1. Bar pattern speeds cover similar ranges in the MaNGA observations and in the TNG50 simulation, differing in the mean by only 15%. Bars in the cosmologically-simulated galaxies are not slow, as the analysis of previous (lower-resolution) simulations suggested. This now gives credence to the use of cosmological simulations when relating the observable properties of bars to the secular evolution of disk galaxies.

2. The $\mathcal{R} = R_{\rm b}/R_{\rm corot}$ values of bars in TNG50 are larger in the average than those in the observed galaxies by 50%. However, we can now attribute this to bars being *shorter*, rather than *slower*. The bar lengths in TNG50 samples the range covered by MaNGA observations, apart from the very longest ones (that are almost exclusively 'observed' in galaxies seen at inclination). Whether the absence of very short bars in MaNGA observed galaxies (classified as 'barred') is an observational classification bias remains to be quantified in our specific work; however, the issue is extensively discussed by Erwin (2018), who concludes that, in SDSS, bars of size below the kpc scale would typically not be resolved.

3. At fixed physical model, we can qualitatively reproduce results of past literature based on previous-generation simulations (Algorry et al. 2017; Peschken & Łokas 2019) by analyzing the output of the same simulated volume of TNG50 but run at lower resolution: this produces bars with larger $\mathcal{R}$ values, lower pattern speeds, and observationally-consistent bar sizes. This indicates that differences between our TNG50-based finding and earlier works can plausibly be attributed to differences in the underlying numerical resolution (as suggested in Algorry et al. 2017), at least partially, i.e. not only because of differences in the underlying galaxy-formation model (as instead previously suggested by Fragkoudi et al. 2020, in AURIGA).

Bar pattern speeds and sizes at $z = 0$ can be a powerful probe of both the physical processes dominating the evolution of the inner galaxies and of the structures that absorbs their angular momentum, including dark matter. But as the present-day bar properties result from their formation properties and their subsequent evolution, it will be important in the future to understand the formation mechanisms of bars in a cosmological context, and use observational support at higher redshift from e.g. JWST to constrain evolutionary models.

## 7. ACKNOWLEDGEMENTS

It is a pleasure to thank Scott Tremaine for sharing essential wisdom in the late stages of this project and for providing insightful comments after a careful read of the manuscript. We further thank Rainer Weinberger and Francesca Fagkoudi for interesting discussions and Benoit Famaey for interesting discussions and comments on the manuscript. We are grateful to Yetli Rosas-Guevara for sharing her catalogs of barred candidates in TNG50 that we have used for comparison to our methods and to Mark Lovell for producing the SubhaloMatchingLagrangian catalog between TNG50-1 and TNG50-2 that we have used for a test in Appendix. We are grateful to Rui Guo for helpful clarifications on how the bar properties were measured in MaNGA galaxies. NF was supported by the Natural Sciences and Engineering Research Council of Canada (NSERC), [funding reference number CITA 490888-16] through the CITA postdoctoral fellowship and acknowledges partial support from a Arts & Sciences Postdoctoral Fellowship at the University of Toronto. JB acknowledges financial support from NSERC (funding reference number RGPIN-2020-04712). The IllustrisTNG simulations were undertaken with compute time awarded by the Gauss Centre for Supercomputing (GCS) under GCS Large-Scale Projects GCS-ILLU and GCS-DWAR on the GCS share of the supercomputer Hazel Hen at the High Performance Computing Center Stuttgart (HLRS), as well as on the machines of the Max Planck Computing and Data Facility (MPCDF) in Garching, Germany. The computations were performed on the ISAAC cluster of the Max Planck Institute for Astronomy at the Rechenzentrum in Garching. This work has made use of the python libraries matplotlib (Hunter 2007), numpy (Harris et al. 2020), scipy (Virtanen et al. 2020) and astropy (Astropy Collaboration et al. 2013).

APPENDIX

## A. DERIVING THE INSTANTANEOUS PATTERN SPEED

We specify how we extract the pattern speeds from individual snapshots in the TNG50 simulation. Since the output cadence between two consecutive snapshots is greater than the typical angular period of a bar by several factors (except for galaxies within subboxes - that we use to verify our results), we have to use single snapshots from the simulation. The method makes use of the velocity field of stars in the bar region and uses the continuity equation to derive the angular speed of the $m = 2$ Fourier mode.

At given radius R, we have the pattern speed of a structure being the time derivative of its Fourier phase $\phi$

$$\Omega = \dot{\phi}, \tag{A1}$$

where the phase is defined as

$$\phi = \frac{1}{m}\arctan(W_s/W_c), \tag{A2}$$

where $W_c$ and $W_s$ are respectively the real and imaginary components of the Fourier transform of the surface density $\int_{-\pi}^{\pi}\Sigma(R,\varphi)e^{im\varphi}d\varphi$, and its time derivative is (taking only partial derivatives)

$$\dot{\phi} = \frac{1}{m}\frac{\dot{W}_s W_c - W_s \dot{W}_c}{W_c^2 + W_s^2}. \tag{A3}$$

The two azimuthal Fourier components $W_c$ and $W_s$ are the real and imaginary parts, and Monte Carlo integrating in logarithmically spaced radial bins (setting the Surface element $dS \propto R^2 dlnR$) we get

$$W_c = \int_{-\pi}^{\pi}\Sigma(R,\varphi)\cos(m\varphi)d\varphi \approx \sum_{i=1}^{N_\star}\frac{m_i}{R_i^2 dlnR}\cos(m\varphi_i) \tag{A4}$$

$$W_s = \int_{-\pi}^{\pi}\Sigma(R,\varphi)\sin(m\varphi)d\varphi \approx \sum_{i=1}^{N_\star}\frac{m_i}{R_i^2 dlnR}\sin(m\varphi_i) \tag{A5}$$

And their time derivatives are

$$\dot{W}_c = \int_{-\pi}^{\pi}\frac{\partial\Sigma}{\partial t}(R,\varphi)\cos(m\varphi)d\varphi \tag{A6}$$

$$\dot{W}_s = \int_{-\pi}^{\pi}\frac{\partial\Sigma}{\partial t}(R,\varphi)\sin(m\varphi)d\varphi \tag{A7}$$

But we cannot measure $\frac{\partial\Sigma}{\partial t}(R,\varphi)$ directly. However, we know the present-day positions and velocities of the particles, which determine the evolution of their spatial distribution (so the surface density). The continuity equation reads, in cylindrical coordinates,

$$\frac{\partial\Sigma}{\partial t}(R,\varphi) = -\frac{1}{R}\frac{\partial}{\partial R}\left(R\Sigma(R,\varphi)v_R\right) - \frac{1}{R}\frac{\partial}{\partial\varphi}\left(\Sigma(R,\varphi)v_\varphi\right), \tag{A8}$$

where the velocities here are the average velocities at given position of the disk (represent the velocity field). We can plug this back in the above equation, and Monte-Carlo integrate the term $\propto \Sigma(R,\varphi)$

$$\dot{W}_c \approx +\int_{-\pi}^{\pi}\left[-\frac{1}{R}\frac{\partial}{\partial R}\left(R\Sigma(R,\varphi)v_R\right) \right. \\ \left. -\frac{1}{R}\frac{\partial}{\partial\varphi}\left(\Sigma(R,\varphi)v_\varphi\right)\right]\cos(m\varphi)d\varphi \tag{A9}$$

$$\dot{W}_s \approx \int_{-\pi}^{\pi}\left[-\frac{1}{R}\frac{\partial}{\partial R}\left(R\Sigma(R,\varphi)v_R\right) \right. \\ \left. -\frac{1}{R}\frac{\partial}{\partial\varphi}\left(\Sigma(R,\varphi)v_\varphi\right)\right]\sin(m\varphi)d\varphi \tag{A10}$$

The azimuthal velocity of particles $\dot\varphi$ can be obtained from their velocities: $\dot\varphi = \frac{v_\varphi}{R}$ in cylindrical coordinates.

The second term of $\dot{W}_c$ in the brackets (the one in $\frac{\partial}{\partial\varphi}$) can be integrated by parts and reduced to a Monte Carlo integral:

$$\begin{aligned}
&\int_{-\pi}^{\pi}\left[-\frac{1}{R}\frac{\partial}{\partial\varphi}\left(\Sigma(R,\varphi)v_\varphi\right)\right]\cos(m\varphi)d\varphi \\
&= \left[\Sigma(R,\varphi)v_\varphi\cos(m\varphi)\right]_0^{2\pi} \\
&\quad -\int_{-\pi}^{\pi}\frac{1}{R}\Sigma(R,\varphi)v_\varphi m\sin(m\phi)d\varphi \\
&= -\int_{-\pi}^{\pi}\frac{1}{R}\Sigma(R,\varphi)v_\varphi m\sin(m\phi)d\varphi \\
&\approx -\sum_{i=1}^{N_\star}\frac{m_i}{R_i^3 dlnR}v_{\varphi,i}m\sin(m\varphi_i)
\end{aligned} \tag{A11}$$

Similarly, the second term of $\dot{W}_s$ in the brackets (the one in $\frac{\partial}{\partial\varphi}$) can be integrated by parts and reduced to a

Monte Carlo integral:

$$\int_{-\pi}^{\pi} \left[ -\frac{1}{R}\frac{\partial}{\partial \varphi}(\Sigma(R,\varphi)v_\varphi) \right] \sin(m\varphi) d\varphi \\ \approx \sum_{i=1}^{N_\star} \frac{m_i}{R_i^3 dlnR} v_{\varphi,i} m \cos(m\varphi_i) \quad (A12)$$

The term in $\frac{\partial}{\partial R}$ can be spelled out and Monte Carlo integrated to avoid azimuthal binning

$$\int_{-\pi}^{\pi} -\frac{1}{R}\frac{\partial}{\partial R}(R\Sigma(R,\varphi)v_R)\cos(m\varphi)d\varphi \\ = -\int_{-\pi}^{\pi} \frac{1}{R}\Sigma(R,\varphi)v_R \cos(n\varphi)d\varphi \\ -\frac{\partial}{\partial R}\int_{-\pi}^{\pi}\Sigma(R,\varphi)v_R \cos(m\varphi)d\varphi \\ \approx -\sum_{i=1}^{N_\star} \frac{m_i}{R_i^3 dlnR} v_{Ri}\cos(m\varphi_i) \\ -\frac{\partial}{\partial R}\sum_{i=1}^{N_\star}\frac{m_i}{R_i^2 dlnR}v_{Ri}\cos(m\varphi_i) \quad (A13)$$

and same for the term in sine. The differential term, involving a radial velocity gradient, is the most noisy term. This is a trade off between having a purely integral form as in Tremaine & Weinberg (1984b) but using only half of the available dimensions (one spatial position and one velocity component), or using them all but having to consider their gradients. We chose the latter.

We have tested (although do not show) the single-snapshot measurement of bar pattern speeds on a system that is present in one of the high-cadence sub boxes of the simulation that has outputs every few Myr, and for which we can measure directly the bar pattern speed by differentiating the bar phase between two snapshots. The method presented above seems to return estimates of pattern speeds that are good to about 10% accuracy.

## B. MATCHING GALAXIES: OTHER STELLAR MASS ESTIMATES

To quantitatively compare the bar properties in simulation and observations in Section 2, we approximate the effects of selection criteria on observations. In MaNGA (Bundy et al. 2015), galaxies were selected mostly based on (photometric) stellar mass estimates and galaxy size estimates (Wake et al. 2017). In Section 2, for each observed galaxy, we find the 5 nearest TNG50 galaxies in the stellar mass - size plane given the stellar mass estimate in Pace et al. (2019). We test here three different stellar mass estimates for the (same) observed sample of barred galaxies: the estimated total stellar mass from photometry, the total mass from Pace et al. (2019) and the fiducial choice of the mass enclosed within the IFU aperture from Pace et al. (2019). We show how the choice of the stellar mass estimate would affect the main conclusion on the distribution of bar properties in Fig. 7. The main results of this work (the pattern speeds in TNG50 cover the range of the observed pattern speeds, and short bars are over-represented) are not affected, although matching the total mass in TNG50 to the photometric mass estimate in the observations produces the largest deviations.

## C. EFFECTS OF NUMERICAL RESOLUTION

### C.1. Matching Subhalos Across Resolution

To isolate the effect of the resolution alone for a given simulation run with identical initial conditions and physical recipes, we match the Subhalos of the TNG50-1 selected galaxies to the Subhalos of the lower-resolution run TNG50-2. To this aim, we use the results from the SubhaloMatchingLagrangian algorithm (Lovell et al. 2014) and proceed to characterize the properties of the bars in the lower-resolution galaxies. 3 galaxies in the TNG50-1 run do not have a match in the TNG50-2, and 24 matched halos do not have a bar in the lower-resolution run, leading to a final sample reduced to 59 galaxies (from the original size of 86 galaxies matched to MaNGA in Section 2.3). Galaxies matched across runs of different resolution have the same total halo mass (black line in the right panel of Fig. 8). At fixed total halo mass, the same galaxies simulated at lower resolution are smaller depending on their mass[5], have a lower stellar mass and longer and slower bars (Fig. 8). At fixed halo mass, the low-resolution galaxies have smaller pattern speed, by a factor almost constant across the whole halo mass range. On average, the low-resolution bars are slower, longer and stronger than the high-resolution ones.

This means that some aspects that are (at least indirectly) independent of the physical recipes, but that are directly linked to the numerical resolution at which a simulation is run, affect the bar properties in such a way that the resolution of the TNG50-2 run is too low to address the question of whether bars have the right properties for a given physical model. The fact that the pattern speeds of the bars are too low in the TNG50-2 run is qualitatively consistent with previous literature findings that bars are too slow in other simulations run at similar resolutions (e.g. Algorry et al. 2017; Peschken & Łokas 2019). However, differently than what we conclude from our analysis, this result has been so far generally interpreted as an over-efficient dynamical friction by the dark matter component of the simulation.

---

[5] Note that on average, for a normal selection of star-forming galaxies, the lower-resolution run produces larger galaxies (as opposed to here, see Pillepich et al. 2019). This seemingly contradictory results comes from (1) a different selection of galaxies and (2) the fact that this selection lies in the mass range where the size difference in star-forming galaxies is smaller across resolutions (Pillepich et al. 2019, B.1).

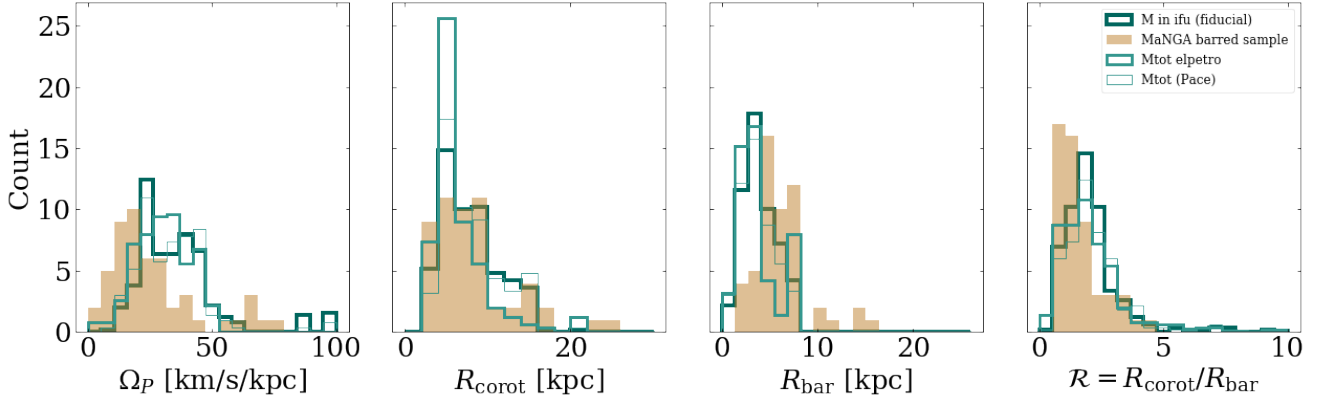

**Figure 7.** Distributions of observed and simulated bar properties using 3 different mass estimates to match the TNG50 galaxies to the observed ones in MaNGA. Using the 'mass in IFU' definition tends to select galaxies of higher mass. As the pattern speed correlates with stellar mass, it shifts the pattern speeds histogram slightly to higher pattern speeds. On the other hand, using the 'M elpetro' definition, which makes different modelling assumptions in the stellar mass measurements, lead to selecting TNG50 galaxies with lower stellar masses. Overall, different stellar mass definitions do not significantly affect our conclusions.

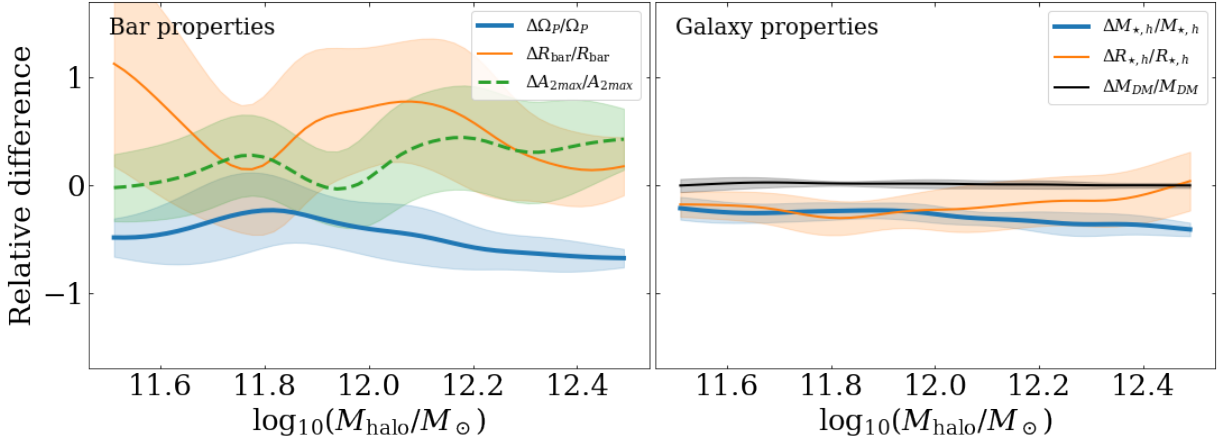

**Figure 8.** Average relative difference of bar (left) and galaxy (right) properties between the high and lower-resolution runs TNG50 and TNG50-2 between galaxies that started with the same initial conditions. On average and across the halo mass range explored here, bars in the lower-resolution run are slower (blue curve below 0 roughly by 40-50%), larger (orange) and slightly stronger (green). The right panel shows that on average, matched galaxies in the lower-resolution run are smaller (see text) and contain less stellar mass in their half mass radius (thick blue). The black line shows that across high- and lower-resolution runs, the halo mass is identical. In all panels, shades show the scatter around the mean relative differences between high and lower-resolution runs.

### C.2. *Dimensional Rescaling of the Lower-Resolution Run*

Resolution affects the distribution of mass within galaxies. Since a different mass distribution may affect the dynamics of galactic systems, we run a qualitative experiment to check whether the bar pattern speeds in low- and high- resolution runs would agree if the mass distribution had been the same. We perform a dimensional rescaling of the lower-resolution run, changing particle galactocentric radii by a factor $r$ and masses by a factor $m$ such that, for each subhalo (subhalo matched between the two runs):

- the bar size in TNG50-2 is the same as in TNG50-1: $r = R_{\text{bar1}}/R_{\text{bar2}}$;

- the mass inside a fractional radius of the bar size $R = fR_b$ is the same in TNG50-2 and in TNG50-1: $m = M(< fR_{bar1})/M(< fR_{bar2})$.

This mass and spatial rescaling requires to change the time units as well, affecting the measurement of the pattern speeds obtained in the TNG50-2 run as $\Omega \propto \sqrt{\frac{m}{r^3}}$, where $m$ and $r$ are the rescaling factors of mass and position determined for each galaxy. After such rescaling, the pattern speeds of the TNG50-2 galaxies aligns with those of the TNG50-1 galaxies (although the scatter also increases), see Fig. 9. To determine which spatial scales are most relevant in the differences between the TNG50-1 and the lower-resolution run TNG50-2, we optimize $f$, the fractional radius of the bar size within which to equate the stellar mass between the two runs. We find (by $\chi^2$ minimization) that $f \approx 0.25$, i.e. scales of a fourth of the bar size, bring pattern speeds to match best between the two runs. These dimensions, probing the inner regions of galaxies, are comparable to the softening lengths in the simulations.

This result implies that at least some of the differences between the two runs purely come from the difference in the inner mass distribution (rather than from evolutionary factors, such as the amount of dark matter that would slow down bars by dynamical friction in different ways in the two runs).

The inner stellar mass content can, to first order, be affected by numerical effect such as the softening length, or the choices of physical recipes applied in the simulation, such as AGN feedback, star formation, stellar feedback, gas physics. As this work aims to focus on the output of such simulations at $z = 0$, we do not explore all these avenues in depth here, but briefly comment on them below.

### C.3. Effects of the gravitational softening

The softening length of the stellar particles in the lower-resolution run of TNG50 nearly doubles compared to that in the high-resolution run, from 288 pc to 576 pc. Taking the softening as a density smoother, (Barnes 2012), we smooth the density of the high-resolution TNG50-1 run with a Kernel set with the softening length of the lower-resolution TNG50-2 run. In Newtonian gravity, this represents the density that generates the gravitational potential that stars would feel if TNG50-1 had the softening length of TNG50-2. Taking the average of the simulated galaxy sample in 3 halo mass bins ($\log_{10}(M_h/M_\odot) < 11.3$, $11.5 < \log_{10}(M_h/M_\odot) < 12$, $\log_{10}(M_h/M_\odot) > 12$), we find that this smoothing procedure matches the central density of the TNG50-2 run at all masses. See Fig. 10 for a summary averaging on all galaxies in our sample.

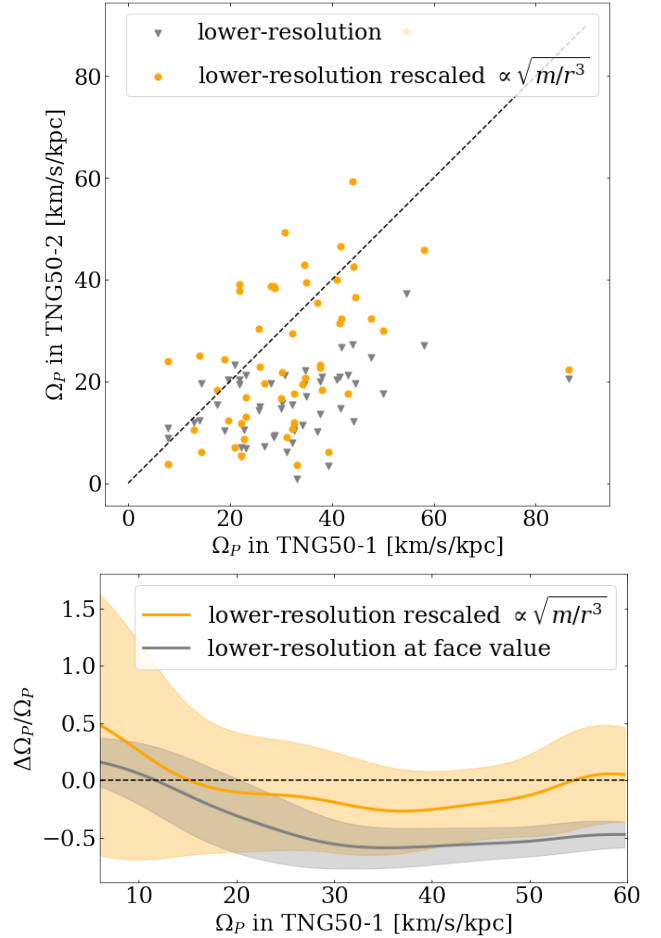

**Figure 9.** A re-scaling of the particles mass and galactocentric radii to homogenize the central stellar mass and bar size induces an alignment of pattern speeds between the lower- and high-resolution runs of the TNG50 simulation.

This means that in the inner regions of the galaxies, the gravitational softening suppresses some of the self-gravity, and stellar particles cannot react to a dense and concentrated inner mass distribution. Their orbits and orbit distribution, which only feel the softened potential, then have a limited inner density. This central smoothing leads to density profiles that cannot 'host' nor 'resolve' short bars.

The bar properties (including its pattern speed) are sensitive to the gravitational force felt by the simulated particles. To test whether the difference in gravitational softening affects the particles' motion by the same amount as the pattern speeds in Fig. 9, we compare the values of the circular rotation speed $\Omega_{circ}(R_b/4)$ at the chararcteristic radius $R_b/4$ of the lower-resolution galaxies, that is shown to be an important scale in Section C.2. The circular rotation speeds are obtained by computing the spherically averaged effective mass en-closed within $R$, from the results obtained to produce Fig. 10, for (1) the high-resolution run at face value, (2) the high-resolution run smoothed with a Kernel with the softening length of the lower-resolution run, and (3) the lower-resolution run at face value. As shown in Fig. 11, softening the high-resolution runs with the lower-resolution softening length brings the circular rotation speeds at $R_b/4$ on the same scale.

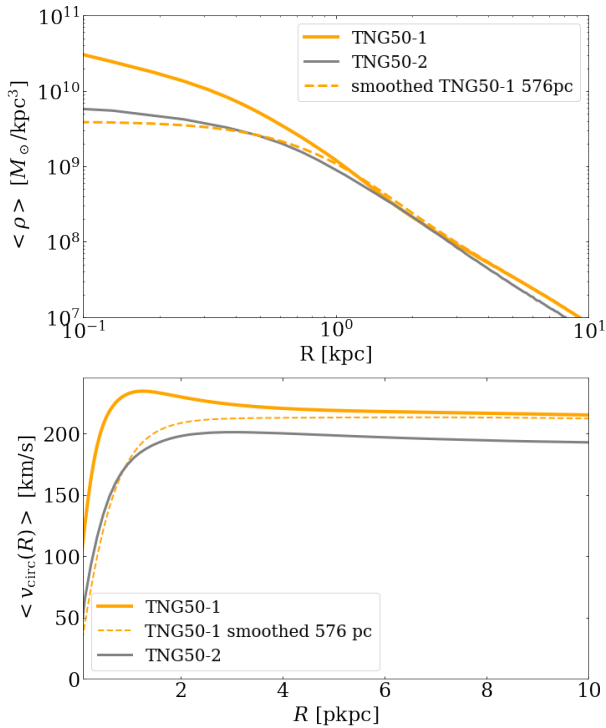

**Figure 10.** Spherically- and sample- averaged mass density in the simulated galaxies used in the analysis (top) and the resulting circular velocity curve (bottom), for both the high-resolution run (solid orange), the lower-resolution run (grey), and the high-resolution run that is smoothed in post-processing with a kernel of the lower-resolution softening length (dashed orange). This brings the central densities of the TNG50-1 and TNG50-2 galaxies on the same scale, especially near $R = 0.5$ kpc (a small fraction of the average bar size). The same conclusions are reached when averaging in bins of stellar mass instead of the whole sample, which we do not show here for clarity.

Therefore, the force softening length could be indirectly responsible for a difference in the inner density profile, which itself can explain much of the differences between the bar properties in the TNG50-1 and TNG50-2 galaxies (Section C.2).

### C.4. *Baryon Dominance*

Fragkoudi et al. (2020) proposes that the increased baryon dominance in the AURIGA zoom-in simulated galaxies compared to galaxies from previous simulations (EAGLE and Illustris) could be responsible for bars with lower $\mathcal{R}$ and that the numerical resolution difference between these simulations is unlikely linked to the differences in $\mathcal{R}$ across simulations. If bars form from gravitational instabilities due to self gravitation, galaxies with a higher baryon dominance should be more prone to bar formation. This should in turn affect (1) the bar fraction, (2) the bar formation time, and (3) the bar evolution. All this may occur because, a bar that formed earlier has more time to exchange angular momentum with other components of its host galaxy; after bar formation, if the dark matter fraction increases the efficiency of dynamical friction and slows down the bars, Fragkoudi et al. (2020) proposes that galaxies that are more dominated by stars are also typically expected to remain faster than those containing more dark matter.

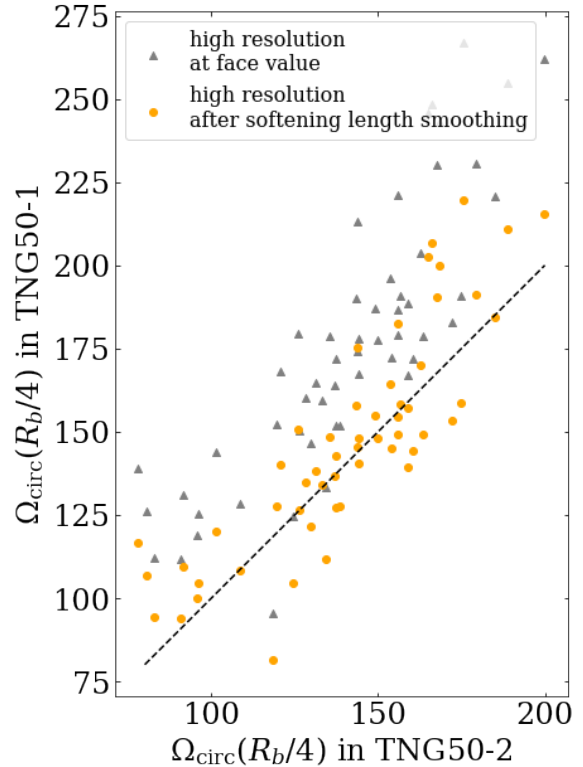

**Figure 11.** Comparing the circular rotation speed $\Omega_{\rm circ}$ at a characteristic radius (a fourth of the bar size measured in the lower-resolution galaxies) obtained in (1) the high-resolution run (grey circles), (2) the high-resolution run smoothed with the softening length of the lower-resolution run (orange triangles), to (3) the lower-resolution run (x axis). The dashed line shows $y = x$ for comparison.

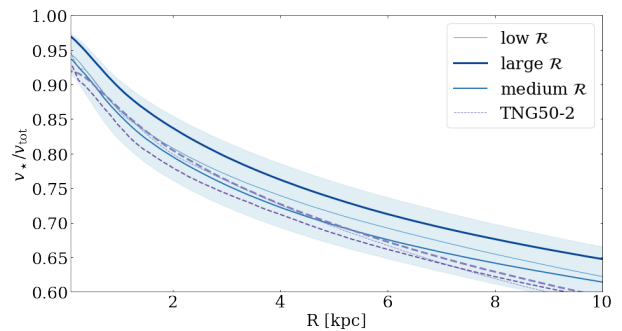

**Figure 12.** Average baryon dominance across the sample of simulated galaxies in bins of $\mathcal{R}$ for the TNG50-1 and TNG50-2 runs. On average, the lower-resolution run (TNG50-2) is less baryon-dominated than the high-resolution one, but the differences across runs are much smaller than the scatter within runs highlighted by the shaded area for TNG50-1. We find no clear trend between the baryon dominance and the $\mathcal{R}$ within a simulation run.

Now, we test this line of arguments within the fixed galaxy-formation model of TNG50. We plot $v_\star/v_{\rm tot}$ as a function of radius, averaged over the simulated galaxies in our TNG50 sample, and find no strong trend between $\mathcal{R}$ and the baryon dominance. However, the galaxies in the lower-resolution run TNG50-2 have a lower baryon fraction than the analogs in the high-resolution run (as already pointed out in Pillepich et al. 2019). The differences we find here with TNG50, at unique physical model, are less notable than those found in Fragkoudi et al. (2020) when comparing different simulations with different models. The increased scatter in such a relation could arise if the galaxies in our simulation have a greater variety in morphology, stellar mass, and accretion history. By construction here, additional components that may play an important role in the bar evolution have here more manoeuvring room due to the wider selection.

### C.5. *Stellar Populations in the Bars*

The differences in stellar populations indicate that the scaling experiments in Sec. C.2 alone cannot explain all the differences between the lower-resolution run and TNG50-1. In star-forming galaxies, bars can grow via the star formation occurring at their ends, providing a set of young stars on cold orbits, more sensitive to non-axisymmetric perturbations. We look for possible differences in the populations of stars that are located in the bar as a function of $\mathcal{R}$ within, and across, simulation runs. We find that in the high-resolution run (TNG50) bars with low $\mathcal{R}$ (dynamically almost filling their corotation radius) have a star formation history on average identical to that of the disk hosting them, implying that part of their formation (or evolution) was recent. Bars with large $\mathcal{R}$ (dynamically short, smaller than their corotation radius) contains stars older than the disk hosting them (but it does not exclude a late bar formation from an old disk). In the lower resolution run, bars tend to have the same star formation history similar to the rest of the disk in all $\mathcal{R}$ bins. This suggests that bars have a different formation history in the high- and lower-resolution runs. Therefore, not all differences can be explained solely by the softening length re-scaling at $z = 0$.

### D. EXAMPLE GALAXIES THAT WOULD PROBABLY NOT BE CLASSIFIED AS BARRED

A few galaxies were classified as 'barred' in our analysis from the mass profile, but they do not look as barred in the light image. If for example the bar is short and weak, it may not be straightforward to identify it in the SDSS-r image: see a couple of examples in Fig. 13.

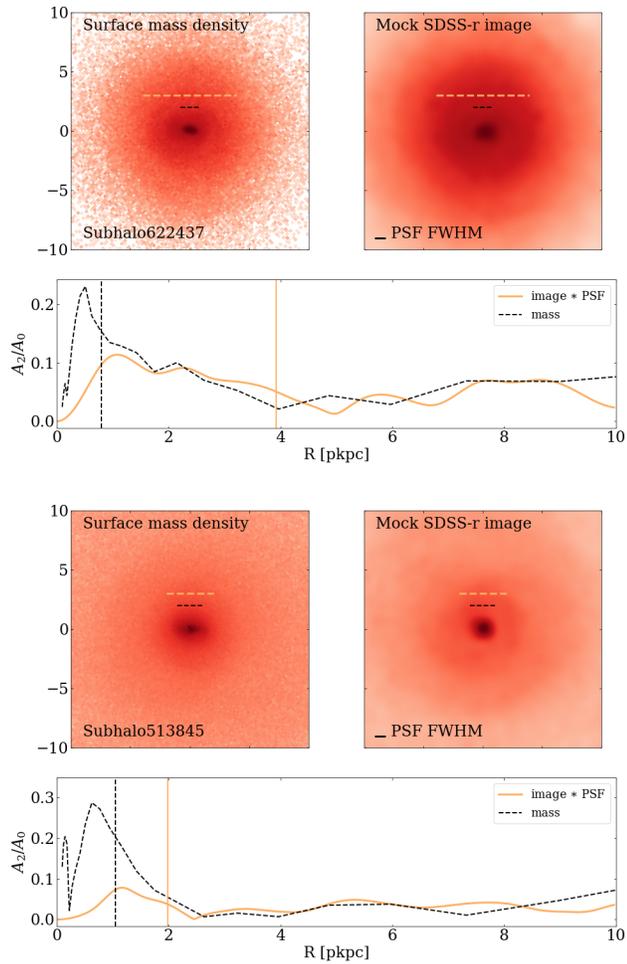

**Figure 13.** Same as Fig. 2 for two TNG50 galaxies whose bar properties differ substantially when inspected based on stellar mass surface density vs. stellar light. In the top, we show a galaxy with an unrealistically large $R_{\rm b,obs}^{TNG50}/R_{\rm b}^{TNG50}$ ratio, from an originally short bar that hardly seems identifiable in the image. In the bottom, we show a TNG50 galaxy with a bar non visible in the mock image. Most of such galaxies have small bars and contribute to increase the $\mathcal{R}$ value of the simulated galaxies, whereas they would not show up in an observed sample.